\documentstyle[12pt]{article}
\input psfig
\def\D0{D\O}
\def\etmisv {\mbox{${\hbox{${\vec E}$\kern-0.6em\lower-.1ex\hbox{/}}}_T$}}
\def\etmis  {\mbox{${\hbox{$E$\kern-0.6em\lower-.1ex\hbox{/}}}_T$ }}

\def\ifmath#1{\relax\ifmmode #1\else $#1$\fi}%

\newcommand{\qqbar}     {\ifmath{\mathrm{q\bar{q}}}}
\newcommand{\ttbar}     {\ifmath{\mathrm{t\bar{t}}}}

\begin{document}
%\setlength{\baselineskip}{3.0ex}
%
%========================================================================
\begin{titlepage}

\begin{center}
\vspace*{3.5cm}
{\large\bf Top quark mass measurements from the Tevatron } \\
\vspace{.5cm}
{\large Rajendran Raja}\\
{\sl Fermilab} \\
\vspace{.2cm}
for \\
\vspace{.2cm}
{\large  The D\O\ and CDF Collaborations} \\
\vspace{.5cm}
\vspace{2.5cm}
\vfil
\begin{abstract}
We review the  measurements of  the top quark mass by the CDF and D\O\
collaborations using Run I data in excess of 100 pb$^{-1}$.  The D\O\
collaboration \cite{d0prl} has recently updated its measurement of the top quark
mass in the  lepton + jets channel. The world average of the top quark mass 
from the CDF \cite{cdf} and D\O\ measurements in the lepton + jets channel now
stands at 175.6 $\pm$ 5.5 GeV/c$^2$.
\end{abstract}

\end{center}
\end{titlepage}

\section{Introduction}
The top quark is of fundamental importance in understanding the standard model
of particle interactions. 
The measurement of the top quark mass and the mass of the $W$ boson can be used
together to constrain the mass of the Higgs particle, an as yet unobserved
particle which is responsible for the generation of
masses in the standard model.
Failure to obtain Higgs mass constraints within reasonable limits
can become the first indication of the incompleteness of the standard model.
This paper reports on the latest measurements of the top quark mass from the 
Tevatron experiments, CDF and D\O. Combining the measurements of the mass from
the lepton + jets channel from the two experiments yields a value for the top
quark mass which has the least error obtained to date. In what follows, we will
report on mass  measurements in three different decay modes of the \ttbar{}
system. The standard model top quark decays into a $b$ quark and a $W$. The
final states of the \ttbar~ system are distinguished by the decay modes of the
two  $W$'s in the event. These are,
\begin{itemize}
\item the dilepton channel, where both the $W$'s decay into a lepton and
a neutrino. Due to the presence of the two neutrinos, this channel is 
underconstrained. Both CDF and D\O\ have differing methods to extract the mass
from this channel.

\item the all jets channel, where both the $W$'s decay into \qqbar{} pairs.
There are no neutrinos  in this channel that can lead to missing transverse
energy.  CDF has reported a measurement  in this channel.

\item the lepton + jets channel, where one $W$ decays into a lepton and neutrino
 and the other decays into \qqbar{} .
 Both CDF and D\O\ have similar methods to extract
the mass information. The D\O\ analysis, using multivariate techniques, is new
and is being reported for the first time at this conference. We will describe
this channel from both experiments more completely than the others, since the
global average mass is being determined using this channel only and also because
the other results have been reported at previous conferences.
\end{itemize}
In all that follows, jet energies have been corrected down to the parton level,
using Monte Carlo models of fragmentation.
In making event selection cuts, however, CDF uses uncorrected jet energies
whereas D\O\ uses corrected energies.
\section{Kinematic fitting}

In the absence of initial or final state gluon radiation, the 6 decay particles
of the \ttbar~ system can be described by 18 variables. Each particle is
described by a three momentum, its energy being  determined given its rest 
mass. 
The hadronic  system that recoils against the \ttbar~ system is
described by its $p_t$, which adds another two variables. There are five
constraints on the model to fit the event, namely that the effective mass of the
two $W$ decay particles has to equal the $W$ mass,  that the effective masses of
the top and anti-top decay products have to be equal to each other and that the
transverse momentum components of the  recoiling hadronic system have to equal
the transverse momentum components of  the \ttbar~ system. This implies that the
theoretical fitting model has 15 free parameters. In the all jets case, all the
final state particles and the transverse momentum components of the recoiling
system are observed, yielding 20 measurements. The system is thus
overconstrained by 5, yielding a 5C fit. In the lepton + jets case, the neutrino
three momentum is unknown, yielding 17 measured variables leading to  a 2C fit.
In the dilepton channel, two neutrinos are missing, yielding 14 measured
variables. This leads to an underconstrained situation ($-1$C fit). If the top
quark mass is specified, one gets a 0C case, and the neutrino solutions can be
determined for each given top quark mass, leading to a likelihood distribution
for the top quark mass for each event. Since in general, one does not know if a
particular jet is the result of a $b$ quark decay, there exist several
permutations of final state particles that must be fitted for the hypothesis in
question. In the all jets case, when both the $b$ quarks are tagged,  there
exist 6 ways of combining the remaining 4 jets into two $W$ decay groups. With
only one $b$ quark tagged, there exist 10 ways of combining the remaining 5 jets
into the two jets associated with the tagged $b$ jet and another three
permutations among the remaining three jets to assign the untagged $b$ jet,
yielding 30 combinations in all. For the lepton + jets  channel, there are 12
combinations in the untagged case and 6 combinations in the tagged case. For the
dilepton channel, there are two combinations for the tagged and untagged case.
The fitting procedure is applied to each combination independently and only
combinations that meet a goodness of fit $\chi^2$ criterion are  kept in the
constrained case. In the unconstrained case, the likelihood distributions from
the combinations are added up and renormalized to obtain the likelihood
distribution for the event. When extra jets are present, the number of
combinations increases rapidly. For this reason, unless otherwise stated, only
the highest $E_T$ jets  are used in the fit, the number of jets being the
minimum required to fulfill the kinematic hypothesis.

\section{The Dilepton channels}

The CDF collaboration \cite {cdf} employs two different techniques in extracting
the mass from the dilepton channels. Using event selection cuts described in the
previous talk \cite{gerdes},  and a cut $H_T>$ 170 GeV, where $H_T$ is the
scalar sum of the $E_T$ of the various objects, CDF obtains 1 candidate in the
$ee$ channel, 1 in the $\mu\mu$ channel and 6 in the $e\mu$ channels. The first
method compares $E_T$ of the found jets with Monte Carlo from different masses. 
This relies on the fact the average $E_T$ of the two $b$ quark jets from  top
quark decay is directly related to the top quark mass. The heavier the top
quark, the more energetic are the $b$ quarks on average. This method yields a
mass of  159 $^{+24}_{-22}(stat)\pm $17(sys) GeV/c$^2$ .

The second method uses the approximate expression $M_{top}^2 = M_W^2 +
2<M_{lb}^2>/(1-<\cos(\theta_{lb})>)$ where  $M_{lb}$ is the lepton $b$ quark
effective mass and $\theta_{lb}$ is the angle between the lepton and the $b$
quark. The quantity $<\cos(\theta_{lb})>$ is obtained from Monte Carlo and a
correspondence function between Monte Carlo and data is used  to calibrate the
result yielding a mass of $162 \pm 21.0 (stat) ^{+7. 1}_{-7.6}(sys)$ GeV/c$^2$.

The D\O\ collaboration uses a variant of the method proposed by Dalitz,
Goldstein \cite{dalitz} and Kondo \cite{kondo} where for each top mass
hypothesis, one tries to obtain solutions for the neutrinos. This results in
0,2 or 4 solutions.  The event four vectors are smeared many times in order to
estimate the probability of neighbouring events fluctuating to give the
observed event.  If one introduces additional information on the QCD production
of the \ttbar{} system, one can weight each solution by a weight proportional
to the product of the structure functions and a decay 
probability\cite{dalitz}. The D\O\ analysis yields a mass \cite{bantly} of 
$158 \pm 24.0(stat) \pm 10(sys)$ GeV/c$^2$.
\section{The All jets channel}
\begin{figure}[h]
\centerline{\psfig{figure=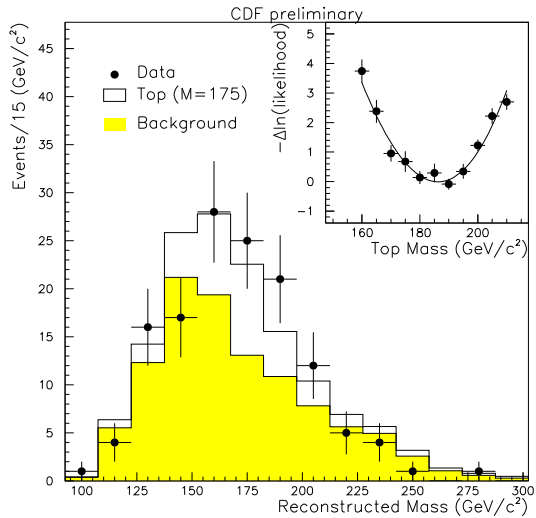,width=3.0in}}
\vskip 0.5cm
\caption{
\label{cdfall}
Likelihood function versus mass for the CDF all jets mass determination
}
\end{figure}

Only the CDF collaboration has reported a mass measurement  in this channel. CDF
performs a three constraint fit in this channel,  (instead of a possible five
constraint fit) by not demanding $p_t$ balance between the recoiling system and
the \ttbar~ system. By demanding $5<N_{jets}<10$ per event and the jet $\Sigma
E_T>200$ GeV, CDF observes 142 events with a $b$ quark tag in the silicon vertex
detector, with a calculated background of 113. This agrees with the rate
expected from top quark production. Performing a kinematical 3C fit yields a
mass \cite{cdf} of $187 \pm 8 (stat) ^{+13}_{-12}(sys)$ GeV/c$^2$ in this
channel. Figure \ref{cdfall} shows the  likelihood function for the CDF mass fit
in this channel.
\section{The lepton + jets channels}
\subsection{CDF results}
The CDF collaboration uses three different techniques to determine the mass from
the lepton + jets channels. The first is largely unchanged from the time of the
top quark discovery \cite{cdfdis}. They require 3 jets with $E_T>$15 GeV and 
pseudo-rapidity $|\eta|<2.0$ and a fourth jet with $E_T>$8 GeV
and $|\eta| <2.4$ . They require at least one jet to be $b$-tagged with the
silicon vertex detector or using a secondary lepton. Thirty four events pass the
selection criteria. All events are treated as a single sample . The estimated
non-top background in this sample is 6.4$^{+2.0}_{-1.5}$ events. This yields a
top quark mass of $175.6\pm5.7(stat)\pm5.9(sys)$GeV/c$^2$.

The second method is called the L$^{**}$ method, where they use in addition the
jet charge and jet tagging probability to enhance the  $\chi^2$ discriminant.
The silicon vertex detector tagging probability is used to assign a jet to be of
primary ($b$ quark) or secondary (W decay) origin. This information is used to
weight each combination. An algorithm is used to estimate the leading quark
charge to discriminate between $b$ and $\bar b$ jets. This method yields a top
quark mass of $174.2\pm5.5(stat)\pm5.3(sys)$GeV/c$^2$.

\begin{figure}[h]
\centerline{\psfig{figure=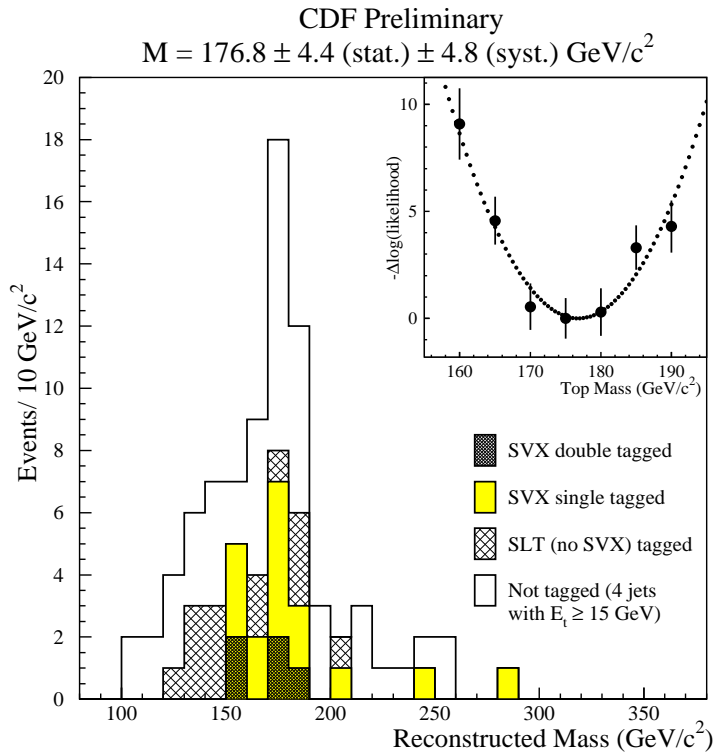,width=4in}}
%\epsfxsize = 16.cm
%\epsffile{t28.EPS}
\caption{
\label{cdfopt}
Reconstructed mass for various sub-samples of the CDF optimized lepton
+ jets fitting. Inset shows the global negative log likelihood function as a
function of the top quark mass}
\end{figure}

The third method used is termed the optimized method, where they divide the
data  into 4 mutually exclusive sub-samples.
\begin{itemize}
\item Events in which a single jet is tagged as a $b$ quark jet using the 
silicon vertex detector (SVX). 
This yields a mass of $176.3\pm8.2(stat)$GeV/c$^2$.
\item Events in which two jets are  tagged as  $b$ quark jets using the 
silicon vertex detector (SVX).
This yields a mass of $174.3\pm7.9(stat)$GeV/c$^2$.
\item Events in which one or more jets are tagged as a $b$ quark 
using an associated lepton and no SVX tag is present.
This yields a mass of $140.0\pm24.1(stat)$GeV/c$^2$.
\item Untagged events.
This yields a mass of $180.9\pm6.4(stat)$GeV/c$^2$.
\end{itemize}
The signal to background in the untagged events is increased by demanding all
jets to have $E_T>15$ GeV. Background analysis is performed for each individual
subsample and separate fits are performed for each subsample.  The results of
the optimized fitting method are shown in figure \ref{cdfopt}. The optimized
method results form the main CDF measurement in this channel, the other two
methods serving as cross checks.
The systematic errors for the optimized method have been worked out in detail
and are shown in table \ref{cdfd0sys}. The  sub-sample likelihoods are combined
into one global likelihood yielding a mass of  $176.8\pm4.4(stat)\pm 4.8(sys)$
GeV/c$^2$ for the optimized method.
\begin{table}[h]
\begin{center}
\begin{tabular}{|c|c||c|c|} 
\hline
\hline
CDF Systematic & Error     & D\O\ systematic & Error   \\
               & GeV/c$^2$ &                 & GeV/c$^2$ \\
\hline\hline
 Soft Gluon + Jet $E_T$ scale & 3.6 & Jet energy scale & 4.0\\
 Monte Carlo event Generators & 1.4 & Generator ISAJET/HERWIG & 3.3\\
                              &     & Generator VECBOS        & 2.6 \\
 Hard Gluon Effects           & 2.2 & Multiple interactions   & 1.2 \\ 
 Kinematic and Likelihood fitting methods & 1.5 & LB/NN difference & 1.4 \\
 b-tagging bias               & 0.4  & Likelihood fit               & 1.0 \\
 Monte Carlo statistics       & 0.8 & Monte Carlo statistics  & 0.8 \\
 Background spectrum          & 0.7 & &\\
\hline
\hline
CDF  Total                        & 4.8 & D\O\ total & 6.2\\
\hline
\end{tabular}
\end{center}
\caption{
\label{cdfd0sys}
Systematic uncertainties for the CDF optimized method and the D\O\
multivariate methods}
\end{table}
\subsection{D\O\ results}
D\O\ uses two independent multivariate techniques to extract the signal.
Multivariate techniques permit the separation of signal from background by
using an appropriate discriminant that is a function of more than one variable.
These techniques are superior to the conventional cuts method prevalent in high
energy physics where signal is separated from background by cutting on single
variables sequentially. When many variables are needed to separate signal from
background, the cuts method results in serious losses of signal, especially in
cases when the amount of signal is small and the signal to background ratio is
small in any given variable.
\subsubsection{Event selection}
D\O\ selects events with electrons(muons) with $E_T>20$ GeV, with $|\eta|<2.0
(1.7)$  and missing transverse momentum  $\etmis >20 (25) $ GeV. D\O\ further
demands $\geq 4$ jets with $E_T>$ 15 GeV and $|\eta_{jet}|<$ 2.0. There is also
a cut on the scalar $E_T$ sum of the $W$ leptonic decay products, $E_T^L \equiv
(E_T^{lepton} + \etmis) > 60$ GeV, with $|\eta_W|<$ 2.0, for events without a
$b$ quark muon tag. Events which have a $b$ tag are selected with  $p^{\mu}_T>$4
GeV, with the muon within $\Delta R \equiv \sqrt{(\Delta \eta^2 + \Delta \phi
^2)}<0.5 $ of a jet. These cuts yield 91 events of which 7 are $b$ tagged. A 2C
top mass fit is performed to these events which yields 77 events with $\chi^2 <$
10, of which 5 are $b$ tagged and $\approx$ 65$\%$ are background. HERWIG Monte
Carlo \cite{herwig} is used to simulate the signal, VECBOS MC  \cite{vecbos} is
used to simulate  the dominant $W$ + multijet background. The $\approx$ 20$\%$
of background events from non-W sources are modeled by multijet data, where one
of the jets fluctuates to a lepton that almost passes the lepton selection.

\subsubsection{Multivariate methods}
D\O\ defines 4 variables to be used in the multivariate analysis, which are so
chosen to enable us to obtain good  signal to background differentiation. These
are

\begin{figure}[h]
\centerline{\psfig{figure=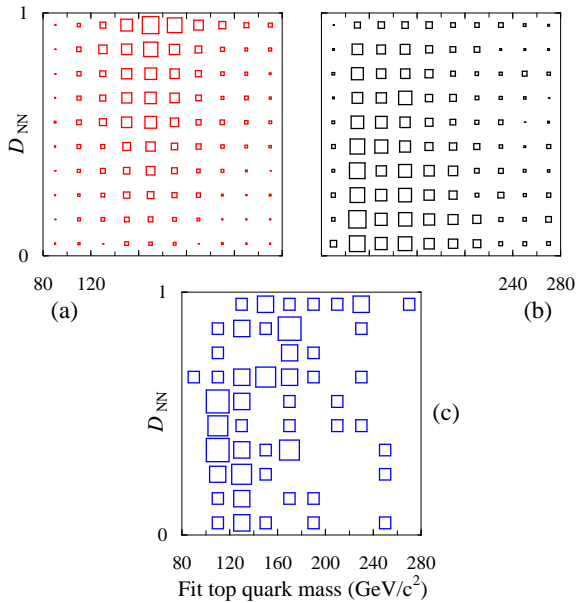,width=3.0in}}
%\epsfxsize = 10.cm
%\epsffile{prl-fig2.eps}
%\vskip 0.1cm
\vskip 0.5cm
\caption{
\label{nncompare}
Events per bin ($\propto$ areas of boxes) {\it vs.}~$D_{\rm NN}$ (ordinate) 
and $m_{\rm fit}$ (abscissa) for (a) expected 172 GeV/c$^2$ top signal, 
(b) expected background, and (c) data.  }
\end{figure}

\begin{itemize}

\item  $x_1 \equiv$ \etmis (Missing $E_T$)
 
\item $x_2 \equiv {\cal A}$, the aplanarity, being defined as 3/2 the least
eigenvalue of the normalized momentum tensor of the jets and the $W$ boson.

\item $x_3 \equiv$ $\frac{H_T - E_T^{jet1}}{H_z}$ where $H_T$ =  $\Sigma E_T$ of
jets and $H_z \equiv \Sigma |E_z|$ of the lepton, neutrino and the jets, $E_z$
being the momentum component of the object along the beam direction.  $x_3$
measures the centrality of the event.

\item $x_4 \equiv$ $\frac{(\Delta R^{min}_{jj})E^{min}_{T}}{E_T^L}$ , where 
$\Delta R^{min}_{jj}$ is the minimum $\Delta R$ of the six pairs of four jets
and $E_T^{min}$ is the smaller jet  $E_T$ from the minimum $\Delta R$ pair.
This variable measures the extent to which the jets are clustered together.
\end{itemize}
Signal events have larger values of the variables $x_1\ldots x_4$ on average
than the background events that fit with the same value of the top quark mass
as the signal.
D\O\ forms two multivariate discriminants $D_{\rm LB}$ and $D_{\rm NN}$ using
these four variables, where LB stands for the ``low bias method" and NN denotes
a three layer feed-forward neural network with 4 input nodes fed by 5 hidden
nodes and 1 output node. In the LB method, they first parametrize  $L_i(x_i)
\equiv s_i(x_i)/b_i(x_i)$ where $s_i$ and $b_i$ are the  signal and background
densities in each variable, integrating over the others. A log likelihood $\ln
{\cal L} \equiv \Sigma_i w_i \ln L_i $ is computed where the weights $w_i$ are
adjusted slightly away from unity to compensate for any correlation  of $\cal
L$ with the fitted top quark mass. $D_{\rm LB}$ is then defined as ${\cal L}/(1
+ {\cal L})$. An event is accepted as a signal event if there exists a $b$
quark tag. It is taken as a background event  if $D_{\rm LB}<$0.43 or if $H_{T}
-E_T^{jet1} <$ 90 GeV. In the neural network approach, the network output
$D_{\rm NN}$ approximates the ratio $s(x)/(s(x)+b(x))$. 
Figure \ref{nncompare} displays the correlation between $D_{\rm NN}$ and the
fitted top quark mass $m_{\rm fit}$ for signal events, background events and
data. Signal peaks at high values of $D_{\rm NN}$ at the generated top quark
mass of 172 GeV/c$^2$ whereas the background peaks at lower values of $m_{\rm
fit}$ and $D_{\rm NN}$.
\begin{figure}[h]
\centerline{\psfig{figure=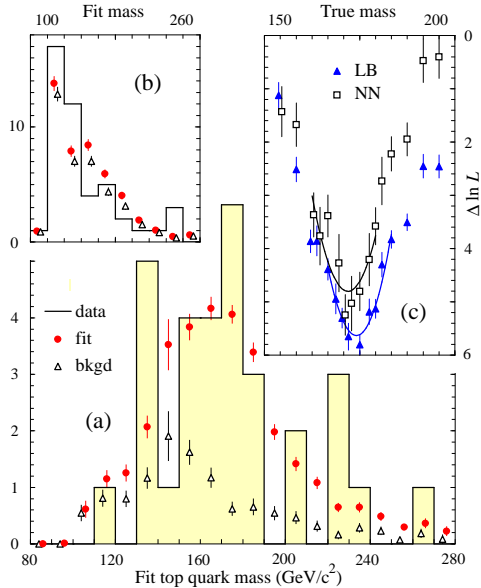,width=2.5in}}
%\epsfxsize = 10.cm
%\epsffile{prl-fig3.eps}
%\vskip 0.1cm
\vskip 0.5cm
\caption{
\label{d0fit}
(a--b) Events per bin {\it vs.}~$m_{\rm fit}$ for events (a) passing or (b) 
failing the LB cut.  Histograms are data, filled circles are the predicted 
mixture of top and background, and open triangles are predicted background 
only.(c) Log likelihood $L$ {\it vs.}~true top quark mass $m_t$ for the LB
(filled  triangles) and NN (open squares) fits, with errors due to finite top
MC  statistics.}
\end{figure}
\begin{figure}[htb]
\vbox{
\centerline{\psfig{figure=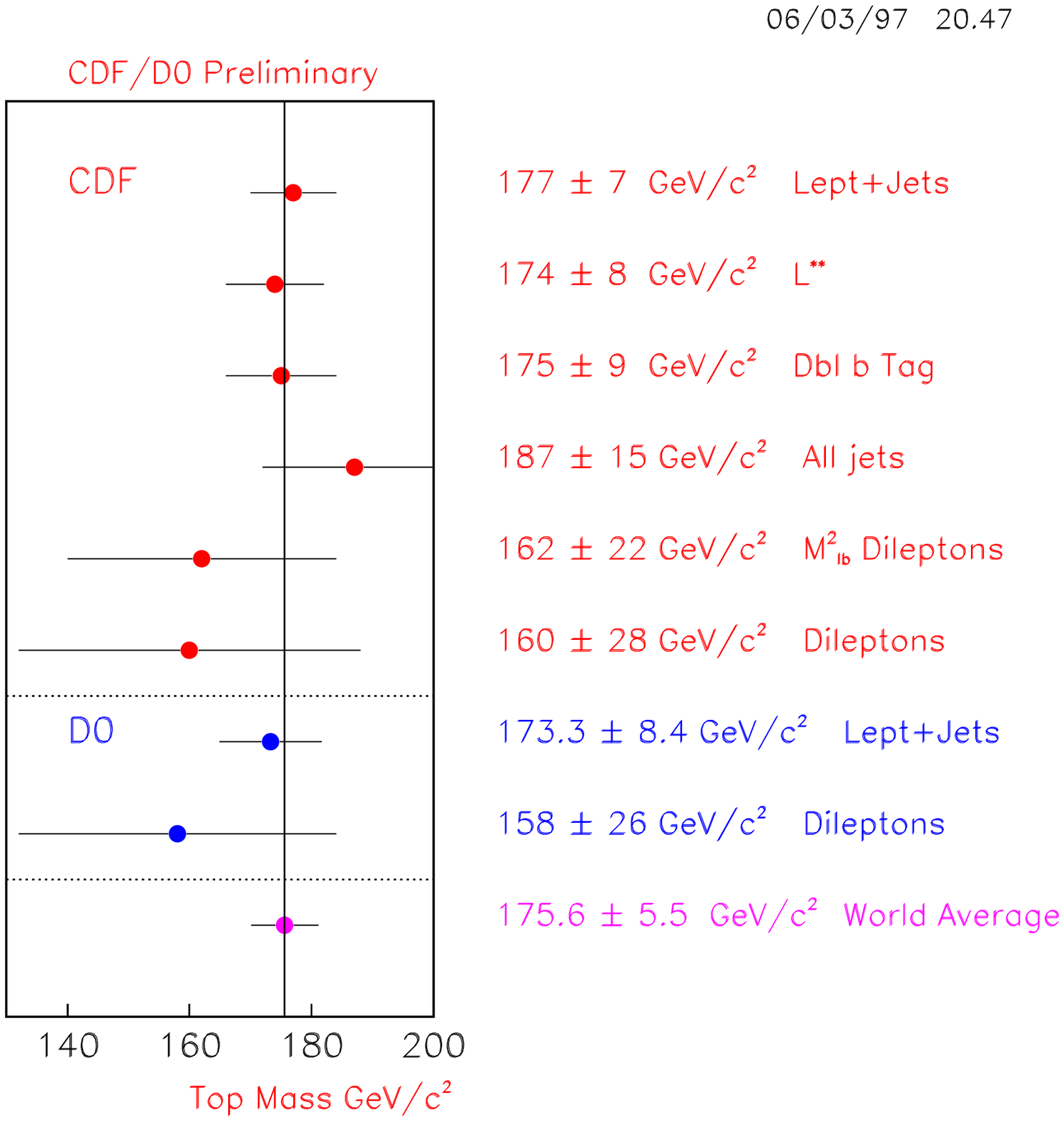,width=4.0in}}
%\vskip 0.1cm
\vskip 0.5cm
\caption{
\label{massall}
Summary of mass measurements and errors from the two experiments using 
various methods. The world average, indicated by the vertical line is obtained
using the lepton + jets channel measurements from CDF and D\O\ }
\centerline{\psfig{figure=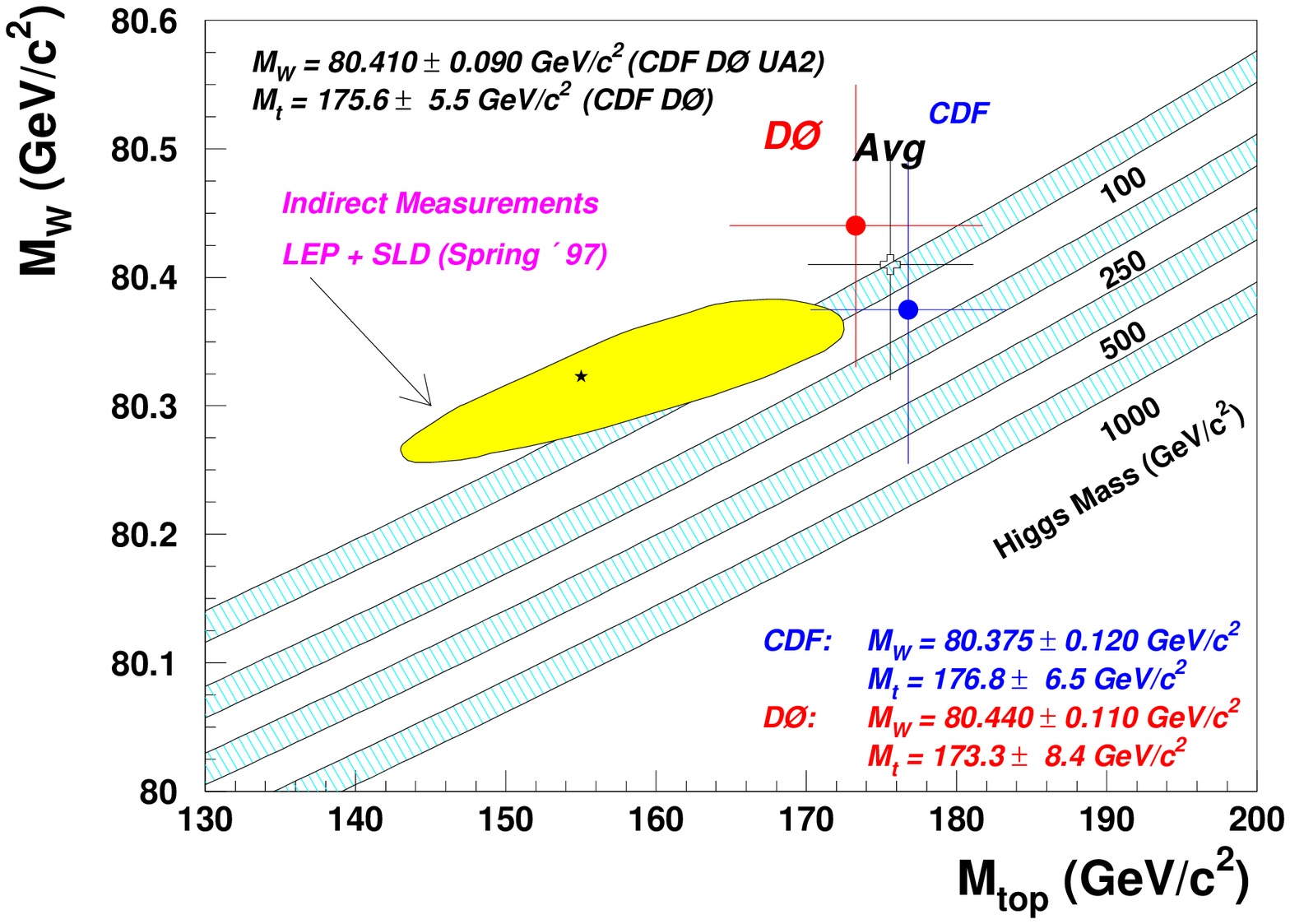,width=4.5in}}
%\vskip 0.1cm
\vskip 0.5cm
\caption{
\label{mwmt}
Comparison of $M_W$ vs $M_{top}$. Contour shows regions allowed by LEP and SLD
data   }
}
\end{figure}

Figure \ref{d0fit} shows the distributions of $m_{\rm fit}$ for data (a)
passing and (b) failing the LB cut. The likelihood analysis  proceeds by
binning the LB(NN) methods in 40(200) bins in $D_{\rm LB/NN}$ $m_{\rm fit}$
space and for each bin maximizing the likelihood L($m_t,n_s,n_b$) assuming
Poissonian statistics and Bayes' theorem\cite{bps}; $n_s,n_b$ are the signal and
background events expected in the data. Figure \ref{d0fit}(c) shows the negative
log likelihood thus  obtained as a function of the top quark mass.  This yields
a top quark mass of $174.0 \pm 5.6(stat)$ GeV/c$^2$ for the LB method and 
$171.3 \pm 6.0(stat)$ GeV/c$^2$ for the NN method.  Table \ref{cdfd0sys} shows
the breakdown of the systematic errors for D\O. The LB and NN methods are
correlated $88\pm4\%$ with each other.  The two results are combined taking into
account these correlations yielding  a top quark mass of $173.3 \pm 5.6(stat)
\pm 6.2 (sys)$ GeV/c$^2$ .

\subsection{Combining CDF and D\O\ results}
 In order to combine the two results, it is necessary to estimate the common
 systematic between the two experiments. These occur in the areas of Monte Carlo
 generators for signal and background, parton fragmentation and  
 luminosity related systematics. While it is certainly not a very well defined
 process, reasonable people can agree that the common systematic error between
 CDF and D\O\ is conservatively in the neighborhood of 3.0 GeV/c$^2$. With this 
 assumption, the CDF optimized method result and the D\O\ multivariate method
 result can be combined to yield a world average top quark mass of 
 $175.6 \pm 5.5$ (stat and sys) GeV/c$^2$. Figure \ref{massall} shows the mass
 measurements from the various channels and their errors from the two
 experiments.
\section{$M_W$ vs $M_{top}$}
Using a world average $W$ boson mass of $80.410 \pm 0.090$ GeV/c$^2$\cite{wm},
and the currently obtained  top quark mass, we obtain the comparison shown in
figure \ref{mwmt}. Within the currently prevalent errors, the standard model is
in good agreement, the data perhaps favoring lower Higgs masses.
\section{Conclusion}
We present the measurements of the top quark mass  and its world average using
lepton + jets channels from the  CDF and D\O\ experiments at the Fermilab
Tevatron. The top quark mass measurements using other decay channels are in
agreement with the world average within errors. A comparison of the top quark
mass and $W$ mass world average with the predictions of the standard model
radiative corrections show no disagreement from what is expected in the standard
model.  Further large improvements in the top quark mass measurement error must
await data from the upgraded  Tevatron and detectors.

\end{document}